\documentclass[useAMS,usenatbib]{mnras}
\usepackage{graphicx,fleqn,fix2col,rotating}
\DeclareGraphicsExtensions{.eps,.ps,.eps.gz,.ps.gz,.eps.Z}
\title[LT follow-up during the first run of aLIGO]{Liverpool Telescope follow-up of candidate electromagnetic counterparts during the first run of Advanced LIGO}

\author[C.M.~Copperwheat et al.]{C.M.~Copperwheat$^{1}$, I.A.~Steele$^{1}$, A.S.~Piascik$^{1}$, D.~Bersier$^{1}$, M.F.~Bode$^{1}$,\newauthor C.A.~Collins$^{1}$, M.J.~Darnley$^{1}$, D.K.~Galloway$^{2,3}$, A.~Gomboc$^{4,5}$, S.~Kobayashi$^{1}$,\newauthor G.P.~Lamb$^{1}$, A.J.~Levan$^{6}$, P.A.~Mazzali$^{1,7}$, C.G.~Mundell$^{8}$, E.~Pian$^{1,9,10}$,\newauthor D.~Pollacco$^{6}$, D.~Steeghs$^{6,2}$, N.R.~Tanvir$^{11}$, K.~Ulaczyk$^{6}$ and K. Wiersema$^{11}$\\\\
$^{1}$ Astrophysics Research Institute, Liverpool John Moores University, IC2, Liverpool Science Park, L3 5RF, UK\\ 
$^{2}$ School of Physics \& Astronomy, Monash University, VIC 3800, Australia\\ 
$^{3}$ Monash Centre for Astrophysics, Monash University, VIC 3800, Australia\\ 
$^{4}$ Department of Physics, Faculty of Mathematics and Physics, University of Ljubljana, Jadranska 19, 1000 Ljubljana, Slovenia\\ 
$^{5}$ University of Nova Gorica, Vipavska 13, 5000 Nova Gorica, Slovenia\\
$^{6}$ Department of Physics, University of Warwick, Coventry, CV4 7AL, UK\\ 
$^{7}$ Max-Planck-Institut f\"ur Astrophysik, Karl-Schwarzschild Strasse 1, D-85748 Garching bei M\"unchen, Germany\\
$^{8}$ University of Bath, Claverton Down, Bath, BA2 7AY, UK\\ 
$^{9}$ INAF, Istituto di Astrofisica Spaziale e Fisica Cosmica, via P. Gobetti 101, 40129 Bologna, Italy\\ 
$^{10}$ Scuola Normale Superiore di Pisa, Piazza Dei Cavalieri 7, 56126 Pisa, Italy\\
$^{11}$ Department of Physics and Astronomy, University of Leicester, University Road, Leicester LE1 7RH, UK\\
}

\date{Received: }

\begin{document}

\def\rchi{{${\chi}_{\nu}^{2}$}}
\newcommand{\Msun} {$M_{\odot}$}
\newcommand{\Mwd} {$M_{wd}$}
\newcommand{\Mbh} {$M_{\bullet}$}
\newcommand{\Lsun} {$L_{\odot}$}
\newcommand{\Rsun} {$R_{\odot}$}
\newcommand{\Zsun} {$Z_{\odot}$}
\newcommand{\Mjup} {$M_{J}$}
\newcommand{\Rjup} {$R_{J}$}
\def\Mdot{\hbox{$\dot M$}}
\def\mdot{\hbox{$\dot m$}}

\newcommand{\ha}{\hbox{$\hbox{H}\alpha$}}
\newcommand{\hb}{\hbox{$\hbox{H}\beta$}}
\newcommand{\hg}{\hbox{$\hbox{H}\gamma$}}
\newcommand{\heii}{\hbox{$\hbox{He\,{\sc ii}\,$\lambda$4686\,\AA}$}}
\newcommand{\hei}{\hbox{$\hbox{He\,{\sc i}\,$\lambda$4472\,\AA}$}}

\maketitle

\begin{abstract} 
The first direct detection of gravitational waves was made in September 2015 with the Advanced LIGO detectors. By prior arrangement, a worldwide collaboration of electromagnetic follow-up observers were notified of candidate gravitational wave events during the first science run, and many facilities were engaged in the search for counterparts. Three alerts were issued to the electromagnetic collaboration over the course of the first science run, which lasted from September 2015 to January 2016. Two of these alerts were associated with the gravitational wave events since named GW150914 and GW151226. In this paper we provide an overview of the Liverpool Telescope contribution to the follow-up campaign over this period. Given the hundreds of square degree uncertainty in the sky position of any gravitational wave event, efficient searching for candidate counterparts required survey telescopes with large ($\sim$degrees) fields-of-view. The role of the Liverpool Telescope was to provide follow-up classification spectroscopy of any candidates. We followed candidates associated with all three alerts, observing $1$, $9$ and $17$ candidates respectively. We classify the majority of the transients we observed as supernovae. No counterparts were identified, which is in line with expectations given that the events were classified as black hole -- black hole mergers. However these searches laid the foundation for similar follow-up campaigns in future gravitational wave detector science runs, in which the detection of neutron star merger events with observable electromagnetic counterparts is much more likely.
\end{abstract}

\section{INTRODUCTION}  
\label{sec:intro}
The Advanced Laser Interferometer Gravitational-wave Observatory (aLIGO: \citealt{2015CQGra..32k5012A}) made the first direct detection of a gravitational wave (GW) signal on 2015 September 14 \citep{2016PhRvL.116f1102A}. The event's waveform was shown to be characteristic of the compact binary coalescence (CBC) of two stellar-mass black holes. This event, since named GW150914, was detected just prior to the beginning of the first official science run (O1), which lasted until January 2016. In \citet{SecondPaper} it was announced that a second black hole CBC was detected during O1. 

Prior to the beginning of this advanced detector era, a worldwide collaboration of astronomers was established with the aim of detecting electromagnetic (EM) counterparts to a GW event. It is expected that binary mergers involving one or more neutron stars (NS--NS or NS--BH) would show a transient EM signature due to energetic outflows from the merger event \citep{2016arXiv160208492A}. The signature might be the short gamma-ray burst `prompt' emission if the observer is within the opening angle of the jet from the merger event (see, e.g. \citealt{2014ARA&A..52...43B}), or more likely the kilonova emission powered by the radioactive decay of heavy nuclei synthesized in the merger ejecta \citep{1998ApJ...507L..59L,2012ApJ...746...48M,2013ApJ...775...18B,2015ApJ...809L...8K}. This kilonova is more isotropic than the prompt emission, although it may contain a somewhat collimated optical/blue component \citep{2015MNRAS.446.1115M,2015ApJ...813....2M,2015MNRAS.450.1777K,2016MNRAS.tmp..894R}. Very close-by supernovae may also produce GW signatures, whose waveforms depend on whether a transitional massive neutron star is formed, or whether a black hole and accretion disc is the primary source of GW emission.

The  aLIGO and Advanced Virgo (AdV) team can identify a GW candidate within their data stream and propagate this event and its inferred sky location to the EM collaboration as quickly as $30$ minutes after detection of the event, although the latency was much higher in the first science run. The main challenge for EM follow-up is the poor localisation of any event. Currently, with only two GW detectors in the network operational, the median uncertainty in the sky position of any detection is of the order of hundreds of square degrees (e.g. \citealt{2014ApJ...795..105S,2014ApJ...789L...5K,2015ApJ...804..114B,2016LRR....19....1A}). This presents two individual problems: the difficulty of searching such a large area for transient sources in a timely fashion, and secondly the difficulty in distinguishing the true counterpart from the long list of possible candidates which will exist in a sky area of this size. 

The Liverpool Telescope (LT; \citealt{2004SPIE.5489..679S}) is one of the telescopes which participated in the O1 campaign. The field-of-view of the LT ($\sim$$10 \times 10$$'$) makes it poorly suited to the search for candidate counterparts, however its diverse instrument suite makes it the ideal tool for the classification of candidates reported to the collaboration by other facilities. Over the course of the campaign, three triggers were issued to the collaboration. Two of these were for the published events GW150914 and GW151226. The third trigger was for a target with the internal designation G194575 \citep{18442}. A more detailed analysis of G194575 yielded a much reduced significance and it was later retracted after 29 days of follow-up by the EM collaboration \citep{18626}. In this paper we report the LT contribution to the follow-up of these three triggers.

The EM collaboration was unsuccessful in identifying a counterpart to either of the GW events in O1. \citet{2016arXiv160203920C} did report a weak and short-lived transient detected by Fermi $\sim$$0.4$s after the detection GW150914, but this was not detected by any other instrument \citep{2016ApJ...820L..36S}. A non-detection of either event is not surprising given that they were both classified as binary black hole mergers from the GW analyses. The sensitivity of the aLIGO detectors in this run were such that the discovery threshold for a binary neutron star merger event extended to $75$Mpc. The increased sensitivity in future runs will mean a much larger volume of space is probed and the discovery of merger events with EM counterparts is significantly more likely.

In Section \ref{sec:strat} we provide a brief description of the LT, the instrumentation used and the strategy employed for follow-up. In Section \ref{sec:events}, we detail the observations obtained in the aftermath of each trigger and where possible provide classifications for the transient sources observed.  In Section \ref{sec:discuss} we reflect on the first observing run and discuss the LT participation in future campaigns.

\section{LIVERPOOL TELESCOPE FOLLOW-UP STRATEGY}  
\label{sec:strat}

The LT is a fully-robotic, optical/near-infrared telescope with a 2-metre clear aperture, located at the Observatorio del Roque de los Muchachos on the Canary Island of La Palma ($28.7624^\circ$N, $17.8792^\circ$W). One of the key strengths of the facility is the diverse instrument suite: currently seven instruments are simultaneously mounted and available for science operations, allowing for a flexible response to transient follow-up. In O1 the primary instrument used was the {\bf SP}ectrograph for the {\bf R}apid {\bf A}cquisition of {\bf T}ransients (SPRAT: \citealt{2014SPIE.9147E..8HP}), a high-throughput, low-resolution spectrograph designed for the rapid follow-up and classification of supernovae and other transients. SPRAT is a long slit spectrograph with a fixed slit width of $1.8''$ and a wavelength range $4000$--$8000$\AA, with a resolving power of $R=350$ at the centre of the spectrum. The slit and grism are deployable and so the instrument itself is used for acquisition, with a spatial pixel scale of $0.44''$/px. The standard acquisition method for SPRAT is an iterative process whereby an acquisition image is taken, a coordinate system is derived using catalogue stars in the image, and a telescope offset is applied. This is then repeated until the target coordinates are on the pixel corresponding to the middle of the slit. The slit is deployed and an image is taken, and then the grism is deployed for the science exposures.

For some candidates the IO:O imager \citep{2014SPIE.9154E..28S} was also used. This is the optical imaging component of the IO (Infrared-Optical) suite of instruments, and has a $10' \times 10'$ field-of-view with an unbinned $0.15 ''/$px pixel scale, and utilises a 12 position filter wheel. The infrared imaging component, IO:I \citep{2016JATIS...2a5002B} has a fixed $H$-band filter and a $6.27' \times 6.27'$ field-of-view with an unbinned $0.184 ''/$px pixel scale. IO:I was not used during the O1 campaign, although it would have been employed had any potential kilonovae been detected, since the emission from such sources is expected to peak at infrared wavelengths \citep{2013Natur.500..547T,2013ApJ...774L..23B,2015MNRAS.450.1777K}.

Given the large positional uncertainty in any reported GW event, it was impractical for the LT to contribute to the transient search component of the EM programme.  However, the LT observatory does include three small robotic `SkyCams': a project aimed at providing simultaneous wide-field observations in parallel with normal LT data taking \citep{2013AN....334..729M,2014SPIE.9152E..2LB}. All three SkyCams use Andor Ikon-M DU934N-BV cameras. One (`SkyCamA') is used as an all-sky monitor, the other two parallel point with the telescope. `SkyCamT' uses a Zeiss Planar T 85mm f/1.4 ZF2 lens to provide a $9^{\circ}$ field-of-view, and 'SkyCamZ' is mounted inside an Orion Optics AG8 (200mm aperture) telescope, providing a $1^{\circ}$ field-of-view. Consideration was given towards using these instruments to contribute to the transient search in O1 as they had done in the initial LIGO and Virgo era \citep{2014ApJS..211....7A}. However, the faintest sources detectable in SkyCamT are about $R$$\sim$$13$--$14$, which is much brighter than the predictions of kilonova models. The limiting magnitude of SkyCamZ is $R$$\sim$$18$, which is potentially useful, however the field-of-view of this instrument compares poorly with other dedicated survey facilities, and the SkyCam archive contains historic data for only $\sim$$15$ per cent of the visible sky, making transient identification difficult. It was therefore decided that our best contribution to O1 would be to follow-up transients detected by other facilities, and in particular provide the rapid spectroscopic classifications which are a core strength of the facility.

A complete log of LT observations taken over the course of O1 is given in Table \ref{tab:obs}. All LT data is delivered to users in a reduced form via an automated pipeline. The reductions include bias subtraction, trimming of the overscan regions, flat fielding, bad pixel masking, sky subtraction and (in the case of spectroscopic data) wavelength calibration. Residual cosmic ray features were removed and the spectra were then flux calibrated. The reduced spectroscopic data were analysed using the Supernova Identification code (SNID: \citealt{2007ApJ...666.1024B}) to determine the quantitative classifications we provide in this paper. For spectra where SNID classification was ambiguous we attempted further clarification using Gelato \citep{2008A&A...488..383H}. Where host galaxy redshift was not available we quote in this paper the SNID estimated redshift.

\begin{table*}
\caption{Log of LT observations taken over the first aLIGO science run. For each object we provide the candidate ID and detection date, time and magnitude, as reported to the EM follow-up collaboration via the GRB Coordinates Network. The remaining columns give the date, time, instrument and exposure times of the LT classification observations. We provide references to the GRB Coordinates Network (GCN) circulars for both the transient discovery and the LT follow-up, and other relevant discovery papers, as follows: $^1$\citet{18362}, $^2$\citet{18370}, $^3$\citet{18371}, $^4$ \citet{2016arXiv160204156S}, $^5$\citet{18497}, $^6$\citet{18549},  $^7$\citet{18553}, $^8$\citet{18573}, $^9$\citet{18762}, $^{10}$\citet{18807}, $^{11}$\citet{18782}, $^{12}$\citet{18791}, $^{13}$\citet{18780}, $^{14}$\citet{18812}, $^{15}$\citet{18832}, $^{16}$\citet{18759}, $^{17}$\citet{18729}, $^{18}$\citet{18742}, $^{19}$Palliyaguri et al. (2016, in prep.)}
\label{tab:obs}
\begin{center}
\begin{tabular}{lllllll}
Candidate ID & Detection        & Detection & LT follow-up     & Instrument & Exposure & References\\
             & date/time        & magnitude & date/time        &            & time (s)    &\\
\hline
\multicolumn{7}{c}{\it GW150914}\\
\multicolumn{7}{c}{\it Event detected 2015-09-14 09:53:51. Alert sent 2015-09-16 05:39}\\
\\      
PS15ccx      & 2015-09-17 13:42 & 19.42 ($z$)     & 2015-09-27 05:31 & SPRAT      & 1800        & 1, 2, 3, 4\\
             &                  &                 & 2016-04-29 20:50 & IO:O      & 150 ($r$)    & \\
\hline        
\multicolumn{7}{c}{\it G194575}\\ 
\multicolumn{7}{c}{\it Event detected 2015-10-22 13:35:44. Alert sent 2015-10-22 20:03:45}\\     
\\       
iPTF-15dkk   & 2015-10-23 05:38 & 19.43     & 2015-10-28 20:34 & SPRAT      & 1500        & 5, 6\\
iPTF-15dkm   & 2015-10-23 03:39	& 18.66     & 2015-10-28 21:41 & SPRAT      & 1500        & 5, 6\\
iPTF-15dkn   & 2015-10-23 05:34 & 19.17     & 2015-10-28 21:09 & SPRAT      & 1500        & 5, 6, 19\\
iPTF-15dld   & 2015-10-23 08:15 & 18.50     & 2015-10-27 23:48 & SPRAT      & 1500        & 5, 19\\
             &                  &           & 2015-11-06 20:54 & SPRAT      & 2100        & 8\\
             &                  &           & 2015-11-06 21:30 & IO:O       & 200 ($u$)   & 8\\
             &                  &           &                  &            & 200 ($g$)   & 8\\
             &                  &           &                  &            & 600 ($r$)   & 8\\
             &                  &           &                  &            & 600 ($z$)   & 8\\             
iPTF-15dln   & 2015-10-23 09:15 & 18.81     & 2015-10-28 00:24 & SPRAT      & 1500        & 5, 6, 19\\
iPTF-15dmk   & 2015-10-23 10:58 & 19.60     & 2015-11-02 23:37 & SPRAT      & 1500        & 5, 7, 19\\
iPTF-15dmn   & 2015-10-23 09:07 & 18.37     & 2015-10-28 22:15 & SPRAT      & 1500        & 5, 6, 19\\
iPTF-15dnh   & 2015-10-23 10:01 & 19.45     & 2015-10-29 01:16 & SPRAT      & 1500        & 5, 6, 19\\
iPTF-15dni   & 2015-10-23 10:23 & 17.72     & 2015-10-28 23:08 & SPRAT      & 1500        & 5, 6, 19\\
             &                  &           & 2015-11-04 20:56 & SPRAT      & 1500        & 8\\
             &                  &           & 2015-11-08 01:25 & IO:O       & 300 ($r$)   & 8\\
             &                  &           &                  &            & 900 (\ha)   & 8\\
\hline    
\multicolumn{7}{c}{\it GW151226}\\      
\multicolumn{7}{c}{\it Event detected 2015-12-26 09:44:37. Alert sent 2015-12-27 17:39:45}\\
\\ 
iPTF-15fed   & 2015-12-28 02:00 & 19.93     & 2016-01-01 20:03 & SPRAT      & 1800        & 9, 10\\
iPTF-15fel   & 2015-12-28 02:27 & 18.96     & 2015-12-30 20:20 & SPRAT      & 1500        & 9, 11, 12\\
iPTF-15fev   & 2015-12-28 02:42 & 17.64     & 2015-12-30 21:55 & SPRAT      & 1500        & 9, 11, 12\\
iPTF-15ffh   & 2015-12-28 03:46 & 18.93     & 2015-12-30 19:04 & SPRAT      & 1500        & 9, 11, 12\\
             &                  &           & 2016-01-02 19:29 & SPRAT      & 1800        & 13\\
iPTF-15ffi   & 2015-12-28 04:06 & 18.71     & 2015-12-30 19:43 & SPRAT      & 1500        & 9, 11, 12\\
iPTF-15ffk   & 2015-12-28 04:06 & 19.19     & 2016-01-01 21:16 & SPRAT      & 1800        & 9, 10\\
iPTF-15ffm   & 2015-12-28 03:47 & 19.81     & 2016-01-03 19:49 & SPRAT      & 1800        & 9, 15\\
iPTF-15ffz   & 2015-12-28 03:44 & 19.65     & 2016-01-01 19:26 & SPRAT      & 1800        & 9, 10\\
iPTF-15fgy   & 2015-12-28 05:47 & 19.52     & 2016-01-01 21:55 & SPRAT      & 1800        & 9, 10\\
iPTF-15fhd   & 2015-12-28 09:42 & 19.70     & 2016-01-03 21:02 & SPRAT      & 1800        & 9, 15\\
iPTF-15fhl   & 2015-12-28 12:42 & 19.09     & 2015-12-31 02:48 & SPRAT      & 1800        & 9, 11, 12, 13, 19\\
             &                  &           & 2016-04-22 22:31 & IO:O       & 150 ($r$)   & \\     
             &                  &           & 2016-04-23 00:49 & SPRAT      & 1800   & \\                     
iPTF-15fhp   & 2015-12-28 04:06 & 19.43     & 2016-01-01 00:58 & SPRAT      & 1800        & 9, 10\\
             &                  &           & 2016-01-02 23:31 & SPRAT      & 1800        & 14\\
             &                  &           & 2016-04-22 21:34 & IO:O       & 150 ($r$)   & \\             
iPTF-15fhq   & 2015-12-28 09:47 & 18.92     & 2015-12-30 21:22 & SPRAT      & 1500        & 9, 11, 12\\
             &                  &           & 2016-04-25 20:50 & SPRAT      & 1500   & \\
             &                  &           & 2016-05-04 21:27 & IO:O       & 150 ($r$)    & \\
iPTF-15fib   & 2015-12-28 05:33 & 19.54     & 2016-01-01 20:39 & SPRAT      & 1800        & 9, 10\\
LSQ15bvw     & 2015-12-29 02:25 & 18.8 ($V$) & 2015-12-29 23:20 & SPRAT     & 1500        & 11, 16\\
MASTER OTJ020906    & 2015-12-27 20:30 & 18.3 (unfiltered) & 2015-12-28 20:59 & IO:O      & 300 ($r$)   & 11, 17\\
                    &                  &                   &                  &           & 900 (\ha)   & \\
UGC 1410 transient  & 2015-12-28 13:42 & 17.3 ($I$)	       & 2015-12-29 19:48 & SPRAT     & 1500        & 11, 18\\
\end{tabular}
\end{center}
\end{table*}

\section{GRAVITATIONAL WAVE EVENTS}  
\label{sec:events}

\subsection{Follow-up of GW150914}  
\label{sec:event1}

\begin{figure}
\centering
\includegraphics[angle=270,width=1.0\columnwidth]{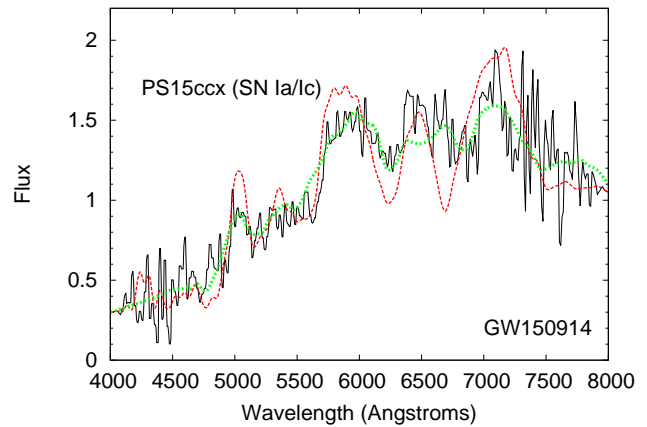}
\caption{Spectrum of PS15ccx obtained during the follow-up campaign of event GW150914. We also plot the best supernova Type Ia (red, dashed) and supernova Type Ic (green, dotted) template fits.} \label{fig:event1} 
\end{figure}

GW150914 was reported in \citet{2016PhRvL.116f1102A} as the signature of the merger of two black holes with masses $36_{-4}^{+5}$ and $29_{-4}^{+4}$\Msun \ at a luminosity distance of $410_{-180}^{+160}$Mpc corresponding to a redshift $z$$\sim$$0.09$. The detection was made at $2015$ September $14$ 09:53UT.

An alert was issued to the EM follow-up collaboration on $2015$ September $16$ \citep{18330}. The alert was not sent in real-time because the detection was made before the formal beginning of O1. A complete overview of the resultant electromagnetic follow-up campaign is given in \citet{2016arXiv160208492A} and \citet{2016arXiv160407864A}. The sky localisation uncertainty region was large, with the northern part close to the Sun at the time of the event making follow-up challenging. However, \citet{18362} reported a number of candidates detected by Pan-STARRS \citep{2010SPIE.7733E..0EK}. A complete account of the Pan-STARRS follow-up of GW150914 is provided in \citet{2016arXiv160204156S}. One of these candidates, PS15ccx, was observed by the LT in evening twilight at $2015$ September $17$ 13:42UT \citep{18370, 18371}. An $1800$ second SPRAT spectrum was obtained, and in \citet{18371} we reported a SNID best-match classification of a supernova Type Ia at age $15$--$23$ days post maximum with $z$$\sim$$0.097$. However, there is a lack of strong Fe lines in the spectrum and so a possible alternative classification is a supernova Type Ic with $z=0.089$. In Figure \ref{fig:event1} we plot the spectrum with both of these template fits.

We observed this target again in April $2016$, seven months after our initial classification, obtaining a deep $r$-band IO:O image. We find no evidence of the transient, suggesting it has faded below the background level of the host galaxy, which we estimate as $r$$\sim$$20.5$ at the object coordinates.

\subsection{Follow-up of G194575}  
\label{sec:event2}

\begin{figure}
\centering
\includegraphics[angle=270,width=1.0\columnwidth]{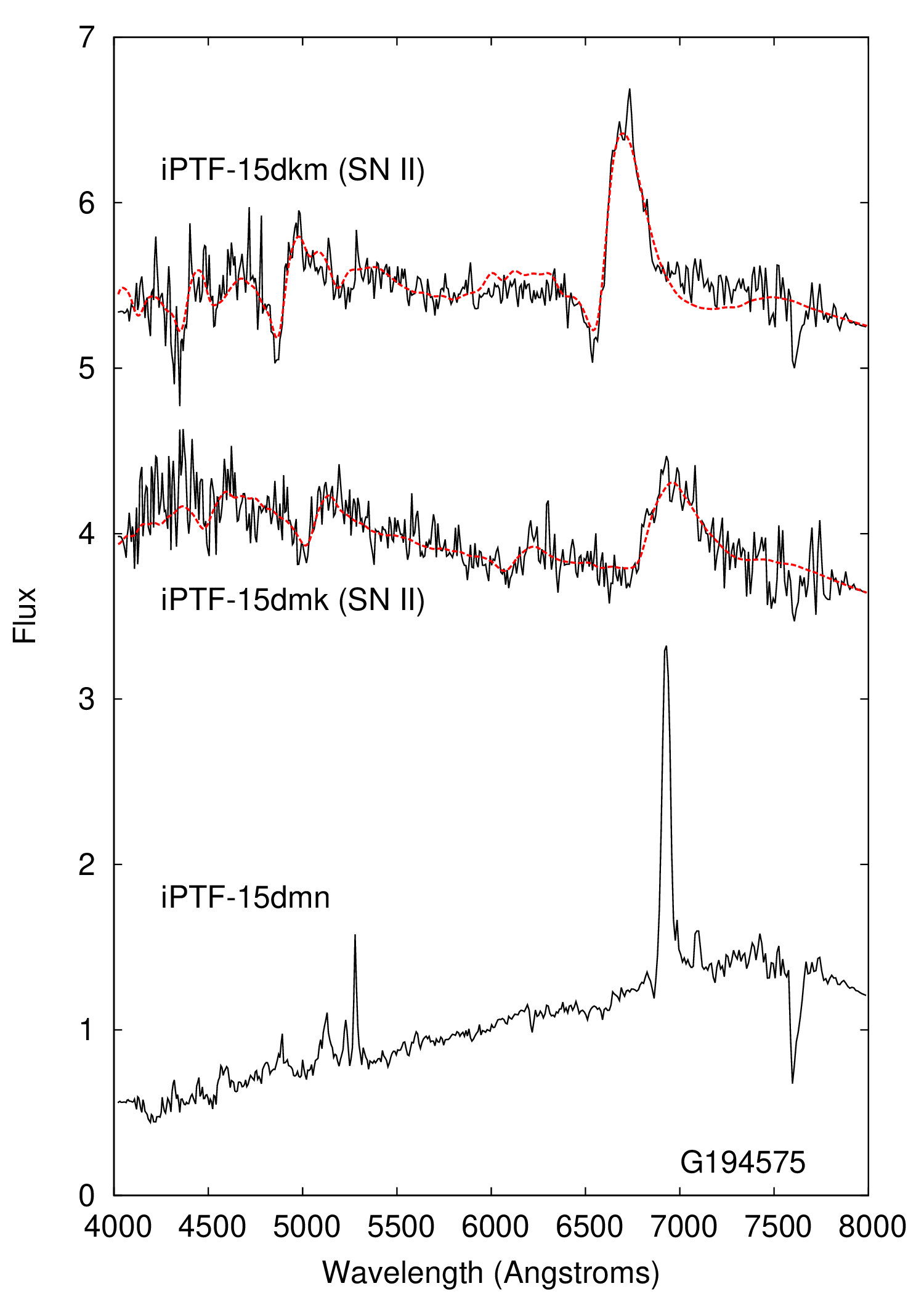}
\caption{Spectra obtained during the follow-up campaign of event G194575. We omit candidates for which our observations showed no evidence of a transient. For objects where a supernova identification is obtained via SNID, we overplot the best template fit (red line, dashed).} \label{fig:event2} 
\end{figure}

An alert was issued on $2015$ October $22$ to the EM follow-up collaboration regarding event G194575 \citep{18442}. The false alarm rate (FAR) of this event was given as $9.65 \times 10^{-8}$Hz, which was above the alert threshold. However, a subsequent offline analysis was reported by the \citet{18626} on $2015$ November $11$ as giving a significantly reduced FAR of $8.19 \times 10^{-6}$Hz, or one in every $1.41$ days, meaning G194575 was determined not likely to be a real event and therefore no longer of interest. However, in the $\sim$$30$ days between these two notifications, a sizeable number of follow-up observations were reported by the collaboration. In this section we detail the LT contribution to this search. 

The LT observations are listed in the second section of Table \ref{tab:obs}. Follow-up data was obtained for nine transients which were reported to the collaboration by \citet{18497}. All of these transients were detected by the intermediate Palomar Transient Factory (iPTF: \citealt{2009PASP..121.1395L}). The iPTF methodology for selecting candidates during the EM follow-up campaign is described in \citet{2016arXiv160208764K}. A summary of our classifications is provided in Table \ref{tab:class2}, and the spectra obtained are shown in Figure \ref{fig:event2}. We comment on individual targets below.

\begin{table*}
\caption{Classifications of transients candidates followed by the LT in response to event G194575. Where we obtain a secure supernova classification we provide a redshift, the time $t$ since peak, and the percentage of matching templates in the SNID database which are consistent with the spectrum.}
\label{tab:class2}
\begin{center}
\begin{tabular}{ll}
Candidate ID & Comments \\
\hline
iPTF-15dkk   & No obvious transient detected. Emission from host galaxy with $z=0.061$\\
iPTF-15dkm   & Supernova Type II, $z=0.03$, $t=+4$ d, $96.5$ per cent template fit\\
iPTF-15dkn   & No obvious transient detected. Emission from host galaxy with $z=0.074$\\
iPTF-15dld   & Some broad emission features, with evidence of contamination by the host galaxy. Consistent with Type Ic supernova.\\
iPTF-15dln   & No obvious transient detected. Spectrum shows host galaxy with $z=0.051$\\
iPTF-15dmk   & Supernova Type II, $z=0.069$, $t=+2$ d, $98.1$ per cent template fit \\
iPTF-15dmn   & Narrow emission lines, consistent with AGN at $z=0.056$\\
iPTF-15dnh   & No obvious transient detected. Emission from host galaxy with $z=0.056$\\
iPTF-15dni   & Weak H-alpha emission with host galaxy absorption at $z=0.020$\\
\hline
\end{tabular}
\end{center}
\end{table*}

\subsubsection{iPTF-15dkk, iPTF-15dkn, iPTF-15dln, iPTF-15dmn, iPTF-15dni, iPTF-15dnh}  
\label{sec:nodetect}

We obtained SPRAT spectra centred on the reported positions of these six targets, and in each case found no evidence of a transient source. The acquisition images indeed show no obvious point source distinct from the host galaxy emission, and the spectra also are apparently dominated by the host galaxy. We estimate the redshifts of the host galaxies from absorption lines in the spectra, and list them in Table \ref{tab:class2}. The spectrum of iPTF-15dmn shows narrow emission lines, from which we classify this host as an AGN with $z=0.056$.

We should emphasise that our non-detections here is not a criticism of the iPTF image subtraction techniques. Typically our non-detection is due to the transient being sufficiently close to its host galaxy nucleus that it cannot be distinguished above the bright galaxy core emission. In other cases the transient may have faded between detection and our follow-up, or our observations might have been taken in poorer conditions.

\subsubsection{iPTF-15dkm, iPTF-15dmk}

We classify iPTF-15dkm as a Type II supernova at approximately $4$ days after explosion at $z$$\sim$$0.03$. A supporting determination was obtained by \citet{18536}, who reported a spectrum obtained $6$ days after ours which they used to classify this object as a Type II supernova at approximately $10$ days after explosion at $z$$\sim$$0.026$.

We classify iPTF-15dmk to also be a Type II supernova, at approximately $2$ days after explosion at $z=0.065$.
 
\subsubsection{iPTF-15dld}

The first follow-up of this transient was reported by \citet{18561}. A re-analysis of these data was then discussed by \citet{18563}, in which the spectrum was found to be consistent with that of a broad line Type Ic supernova, close to maximum light, at a redshift of $z=0.046$. The authors identified this as a source of particular interest, albeit one unrelated to any GW event, since the spectrum resembles that of SN 2006aj \citep{2006Natur.442.1008C,2006Natur.442.1011P,2006Natur.442.1018M}, a rare event likely associated with jet-like core collapse. Further optical observations were reported in \citet{18566}, and this object was also observed with Swift, which did not detect an X-ray counterpart \citep{18569,Evans16}.

The unusual nature of this event prompted us to obtain multiple epochs of spectroscopy supplemented by multi-wavelength photometry. These data will be published in a forthcoming detailed study of this object (Pian et al., 2016, in prep).

\subsection{Follow-up of GW151226}  
\label{sec:event3}

\begin{figure*}
\centering
\includegraphics[angle=270,width=1.0\textwidth]{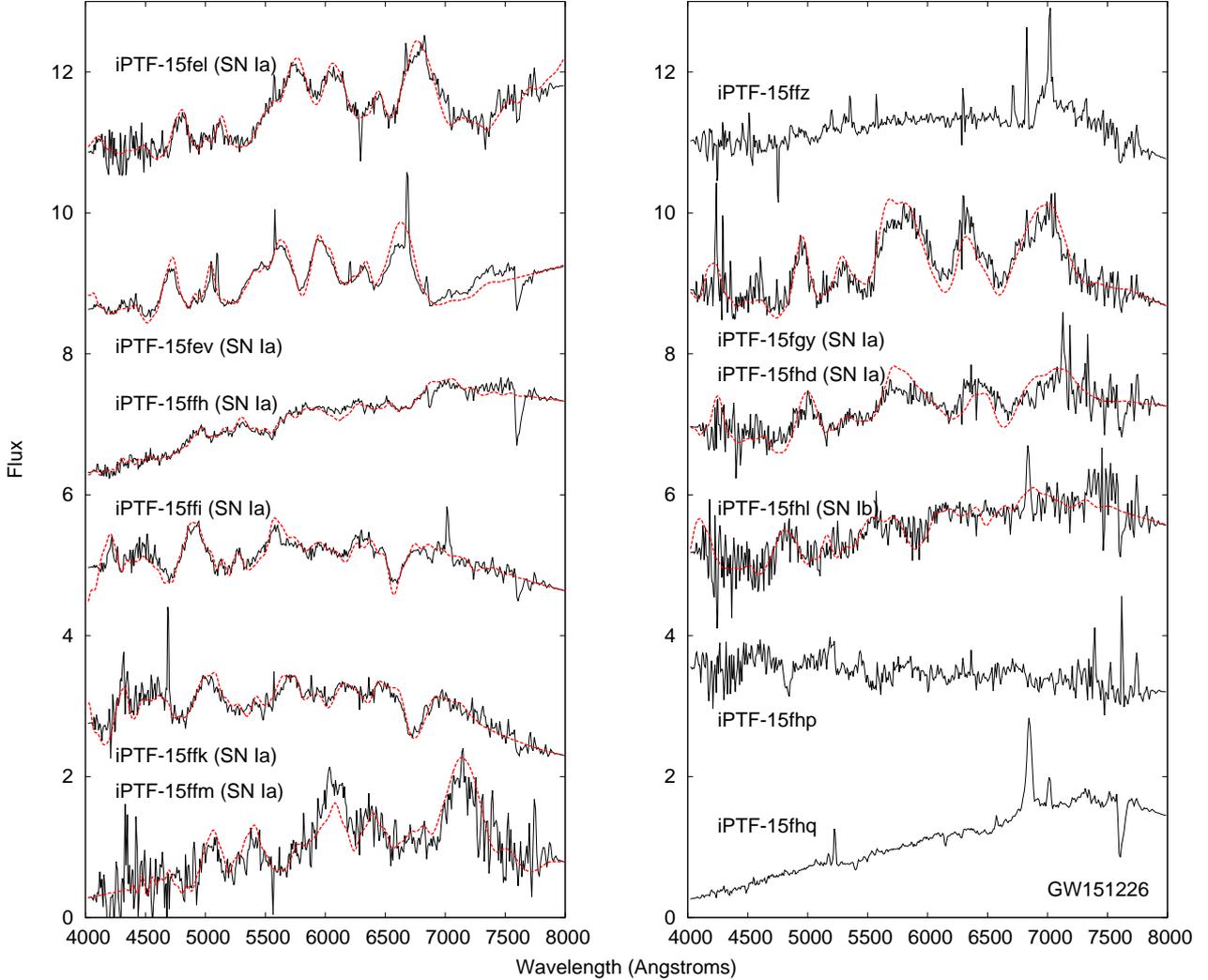}
\caption{Spectra obtained during the follow-up campaign of event GW151226. We omit candidates for which our observations showed no evidence of a transient. For objects where a supernova identification is obtained via SNID, we overplot the best template fit (red line, dashed).} \label{fig:event3} 
\end{figure*}

GW151226 was reported in \citet{SecondPaper} as a CBC of two black holes with masses $14.2^{+8.3}_{-3.7}$ and $7.5^{+2.3}_{-2.3}$\Msun, at a distance of $440_{-190}^{+180}$Mpc ($z$$\sim$$0.09$). The gravitational wave signal was detected at $2015$ December $26$ 03:39UT, and an alert was issued to the EM follow-up collaboration at $2015$ December $27$ 17:40UT, identifying the event as a CBC containing at least one black hole \citep{18728}. The localisation region for this source was well placed for follow-up from La Palma, and we took classification data for $17$ candidates over the course of the week following the alert. The majority of the sources we followed were originally detected by iPTF and reported to the collaboration by \citet{18762}. We were not able to observe every transient in this report, and so prioritised targets without existing spectroscopic classifications with a discovery magnitude less than $20$. A complete account of the iPTF follow-up of GW151226 will be presented in Kasliwal et al. (2016, in prep). A summary of all of the classifications we made is provided in Table \ref{tab:class3}, and the spectra obtained are shown in Figure \ref{fig:event3}. We comment on individual targets below.

\begin{table*}
\caption{Classifications of transients candidates followed by the LT in response to event GW151226. Where we obtain a secure supernova classification we provide a redshift, the time $t$ since peak, and the percentage of matching templates in the SNID database which are consistent with the spectrum.}
\label{tab:class3}
\begin{center}
\begin{tabular}{ll}
Candidate ID & Comments \\
\hline
iPTF-15fed   & No transient detected to limiting magnitude of $R$$\sim$$19.1$\\
iPTF-15fel   & Supernova Type Ia, $z=0.038$, $t=+40$ d, $97.7$ per cent template fit\\
iPTF-15fev   & Supernova Type Ia, $z=0.023$, $t=+50$ d, $94.7$ per cent template fit\\
iPTF-15ffh   & Possible supernova Type Ia, $z=0.061$ $t=+15$d\\
iPTF-15ffi   & Supernova Type Ia, $z=0.085$, $t=+3$ d, $89.1$ per cent template fit\\
iPTF-15ffk   & Supernova Type Ia, $z=0.102$, $t=+5$ d\\
iPTF-15ffm   & Supernova Type Ia, $z=0.094$, $t=+36$ d\\
iPTF-15ffz   & Emission lines consistent with AGN at $z$$\sim$$0.07$\\
iPTF-15fgy   & Supernova Type Ia, $z=0.076$, $t=+20$ d, $84.7$ per cent template fit\\
iPTF-15fhd   & Possible supernova Type Ia, $z=0.091$, $t=+11$ d\\
iPTF-15fhl   & Possible supernova Type Ib, $z=0.043$, $t=+18$ d\\
iPTF-15fhp   & Possible supernova Type Ic, $z=0.129$, $t=+1$ d\\
iPTF-15fhq   & Narrow emission lines, consistent with AGN at $z=0.043$\\
iPTF-15fib   & Slow moving asteroid \\
LSQ15bvw     & No transient detected to limiting magnitude R$\sim$19.5\\
MASTER OTJ020906    &No transient detected to limiting magnitude R$\sim$20\\ 
UGC 1410 transient  &No transient detected. ID'd as minor planet 2 606 Odessa \citep{18762,18822}\\
\hline
\end{tabular}
\end{center}
\end{table*}

\subsubsection{iPTF-15fed, iPTF-15fib, LSQ15bvw, MASTER OTJ020906, UGC 1410 transient}

For these five sources we found no evidence of a transient source in either our SPRAT acquisition image or spectrum. As noted in Section \ref{sec:nodetect}, this could be for a number of reasons: contamination from the host galaxy, a fast fading transient, or poor weather conditions at the time of our follow-up observations. iPTF-15fed in particular was originally reported \citep{18762} with a magnitude that was fainter than the estimated limiting magnitude of our acquisition frames. We obtained $r$-band and \ha \ photometry of MASTER OTJ020906 \citep{18729}, using an \ha \ filter redshifted to a central wavelength of $675.5$nm to match the host galaxy. We did not detect a transient, and observations by \citet{18775} on the subsequent night confirmed this source had already faded below its host galaxy level.

Two of these non-detections were due to the source being a solar system object. iPTF-15fib was identified by iPTF as a slow moving asteroid, and was erroneously included in \citet{18762} as a candidate. \citet{18742} reported a transient in the galaxy UGC 1410. We did not detect a transient in our follow-up data, and this object was subsequently identified as the minor planet 2 606 Odessa \citep{18762,18822}.

\subsubsection{iPTF-15fel, iPTF-15fev, iPTF-15ffi, iPTF-15ffk, iPTF-15ffm, iPTF-15fgy}  

These targets were all classified by SNID as Type Ia supernovae. In Table \ref{tab:class3} we list redshifts and the times in days relative to the time of maximum light.

\subsubsection{iPTF-15ffh, iPTF-15fhd}

We obtained two epochs of observation for iPTF-15ffh. In the first epoch the data were of poor quality and we could not discern any emission from a transient source over that of the host galaxy. In the second visit we obtained a very red spectrum from which we tentatively classify it as a Type Ia supernova using SNID. We also classify iPTF-15fhd as a Type Ia, although again the quality of the data is such that we consider this less secure than some of our other classifications.

\subsubsection{iPTF-15ffz, iPTF-15fhq}

These sources show narrow emission lines, and we classified them as AGN. We observed iPTF-15fhq again in April/May $2016$, four months after our initial observation, and obtained a SPRAT spectrum and a deep $r$-band IO:O image. We find no significant differences in the spectrum or target brightness compared to our first visit, which supports this identification. 

\subsubsection{iPTF-15fhl}

In \citet{18791} we reported that the spectrum of this source shows an emission line on top of the host galaxy spectrum. This is at a wavelength consistent with \ha \ at a redshift of $z=0.044$, which is consistent with the redshift provided by \citet{18762}. A further analysis with SNID identified the transient as a possible supernova Type Ib with $z=0.043$ at $18$ days after peak, although the spectrum suffers from significant contamination from the host galaxy. We  observed this target again in April $2016$, nearly four months after our initial classification, obtaining a deep $r$-band IO:O image and a SPRAT spectrum. The image shows no evidence of the transient, suggesting it has faded below the background level of the host galaxy, which we estimate as $r$$\sim$$20.7$ at the object coordinates. The spectrum shows no emission other than that of the host galaxy, with the \ha \ emission line still present.

\subsubsection{iPTF-15fhp}

We obtained two epochs of observation for this target. The first epoch of data were of poor quality and the best classification was a Type Ia supernova at $7$ days before peak. A second epoch was obtained two nights later and the acquisition image showed that the transient had not brightened, confirming this initial classification to be erroneous. The data obtained for this second epoch were still too noisy for reliable classification, but is consistent with a Type Ic supernova at one day past peak. We observed this target again in April $2016$, nearly four months after our initial classification, obtaining a deep $r$-band IO:O image. We find no evidence of the transient, suggesting it has faded below the background level of the host galaxy, which we estimate as $r$$\sim$$21.5$ at the object coordinates.

\section{DISCUSSION}  
\label{sec:discuss}

\subsection{The challenges of counterpart classification}

The detection of a counterpart to a GW event is a challenging task due to the large uncertainty in the sky position of any detection. This is particularly true in the current era, in which only two GW detectors are operational. However, the capabilities of large survey facilities, and successes such as the $71$~deg$^2$ search for the afterglow to GRB~130702A \citep{2013ApJ...776L..34S,2015ApJ...806...52S}, demonstrate that the problem is not insurmountable. The task is complicated by the fact that a sky area of the order of $100$~deg$^2$ contains many candidate counterparts. Given that the EM signature (rise/fade timescales, colours, etc.) of the true counterpart to a GW event is currently somewhat uncertain, it is difficult to eliminate candidates based on a single detection of a new source alone. Multiple follow-up epochs and complementary bands can distinguish candidates on the basis of their photometric properties, but follow-up spectroscopy is in particular a powerful tool. The transient detection capabilities of modern optical surveys are already such that the available classification facilities struggle to provide comprehensive follow-up in a timely fashion. This is a significant challenge for the successful identification of a GW EM counterpart. This problem will become more serious when the new dedicated survey facilities such as the Gravitational-wave Optical Transient Observer (GOTO\footnote{http://www.goto-observatory.org/}) and BlackGEM \citep{2015ASPC..496..254B} become operational, and existing transient facilities are upgraded. For example, iPTF will be upgraded to the Zwicky Transient Facility within the next year \citep{2014htu..conf...27B}. This will increase the field-of-view of each individual exposure from $7.26$deg$^2$ to $47$deg$^2$, and this coupled with reduced exposure and overhead times results in nearly a $15\times$ improvement in the relative areal survey rate. The LSST era \citep{2009arXiv0912.0201L}, in which the entire Southern sky will be imaged to a depth of $r$$\sim$$24$ every few nights, will present an even greater challenge. Comprehensive exploitation capacity will be required to maximise the time domain science potential of the next generation of surveys. 

For illustration, a census of GRB Coordinates Network (GCN) circulars during the follow-up campaign of GW151226 shows $77$ candidate counterparts discovered by optical surveys and shared with the EM follow-up collaboration. $37$ of these: just under $50$ per cent, were reported to have a firm classification. A further $18$ sources received a slightly more tentative classification based on the survey photometry alone: for example the eventual accumulation of multiple epochs of observation suggesting the target is a flare star, or a dwarf nova. There were also a small number of cases (such as the observations of MASTER~OTJ020906 reported by ourselves and other groups) in which the transient had already faded below its host galaxy level by the time of classification observations. This leaves $19$ candidate counterparts detected by optical surveys and reported to the collaboration, but for which no follow-up classification was attempted. This is a significant fraction of the total number of reported candidates, and demonstrates the `classification gap'. 

The LT follow-up programme described in this paper obtained classification data for $17$ of the $77$ candidate counterparts reported during the GW151226 campaign, which represents a significant contribution to the classification effort. A low-resolution spectrograph such as SPRAT is a powerful tool for follow-up programmes of this nature, and its sensitivity is well matched to the transient discovery space of the iPTF and Pan-STARRS survey facilities, which were the two main contributors of candidates ($20$ in one night of iPTF observations, \citealt{18762}; $44$ over three nights of Pan-STARRS observations, \citealt{18786}). 

The majority of the transients we successfully classified were supernovae, with redshifts in the range $0.022$--$0.129$. For comparison, the redshifts of both GW150914 and GW151226 were determined to be $0.09^{+0.03}_{-0.04}$ 
 \citep{2016PhRvL.116f1102A,SecondPaper}, and at final design sensitivity the advanced detector network will be sensitive to binary neutron star mergers at redshifts of up to $0.045$ \citep{2014ApJ...795..105S,2016LRR....19....1A}. As previously noted the EM signature of the counterpart is not currently clear: different kilonova models produce estimates which span a range of magnitudes (e.g. \citealt{2015MNRAS.450.1777K}). It may be that successful follow-up of a GW counterpart requires a telescope of much larger aperture than the LT. However, we demonstrate here that 2-metre class telescopes are very capable of eliminating and classifying unrelated candidates at ranges comparable to likely GW detections. This will be increasingly important in future aLIGO and AdV runs, since future alerts will include a distance estimate when the candidate is detected via a CBC pipeline. The GW distance can generally be estimated with about a 30 per cent fractional uncertainty \citep{2015ApJ...804..114B,2016arXiv160504242S}. The natural result of this will be targeted follow-up of transients with host galaxies which are consistent with these distance estimates.

\subsection{LT follow-up strategy in future observing runs}

The LT follow-up strategy employed during the O1 generally worked well, although we plan to adopt two changes in future campaigns. First, in O1 we focused on spectroscopic classification of candidates, however we believe it will be worthwhile to also photometrically monitor any candidates which cannot be immediately ruled out as a GW counterpart. This is motivated by alternative theoretical propositions for the counterpart such as \citet{2016arXiv160502769L}, which imply a featureless and unrevealing optical spectrum. Of course, infrared monitoring would still be our primary follow-up methodology in the event of the discovery of a likely kilonova source.

Secondly, in a number of cases we did not detect a transient in our data and we conclude they had faded below the level of its host galaxy by the time of our observations. This is perhaps in part due to our strategy of picking targets from GCN circulars reported to the EM collaboration, meaning that our classification observations typically took place at least a day after the transient discovery. While this response time is generally adequate for supernova follow-up it is not clear if this will be sufficient to detect the signature of a GW counterpart, and this strategy does not play to the rapid response capability which is the core strength of robotic telescopes. We have consequently implemented new observing modes in the telescope software, enabling users to communicate directly with the robotic scheduler via a command line interface rather than the usual web-based tool. This capability has been available for some years for IO:O, and in 2016 we have added this mode for IO:I and SPRAT. The key benefit of this alternative interface is that it makes it straightforward for other robotic facilities to communicate directly with the telescope scheduling software. This provides the prospect of a fully automated follow-up chain, whereby survey telescopes can inject targets directly into the LT observing queue for spectroscopic or infrared follow-up within minutes of transient discovery, and without human intervention. We will be working with other EM follow-up groups in future observing runs to realise this goal. Organised collaborations such as the GROWTH network (Global Relay of Observatories Watching Transients Happen\footnote{http://growth.caltech.edu}) have the potential to be particularly fruitful. New facilities such as GOTO (also located at the Observatorio del Roque de los Muchachos) are also well matched to the LT, probing the transient sky to a depth that is accessible for SPRAT follow-up. 

 The network of GW detectors consisting of two LIGOs, Virgo, KAGRA and possibly a third LIGO in India is expected to be complete by $\ga$$2022$. In the next decade detection and follow-up of candidates is expected to be one of the core science goals of the combined LT transient follow-up facility proposed for the Observatorio del Roque de los Muchachos \citep{2015ExA....39..119C}. This will consist of two telescopes. A new $4$-metre robotic telescope, with the working title Liverpool Telescope 2, which will focus on optical/infrared spectroscopic classification and follow-up; and the existing LT, with its current instrument suite replaced with a prime focus $2 \times 2$ degree imaging camera, which will be used for candidate detection.

\section{CONCLUSIONS}  
\label{sec:concs}

We have presented a summary of the LT observations obtained in support of the first aLIGO science run. The LT was part of a worldwide collaboration of EM facilities which responded to three triggers over the course of the run with the aim of detecting the EM counterpart to a gravitational wave event. The collaboration was unsuccessful in identifying a counterpart. However, while two of the triggers were associated with statistically significant gravitational wave events, those events have been interpreted as resulting from the merger of a binary pair of black holes, from which no EM counterpart is expected. Future gravitational wave detector observing runs will probe a larger region of space, increasing the likelihood of the detection of an `EM bright' merger event involving at least one neutron star component, and the first observing run can be viewed as a preparatory step towards those future campaigns.

The key challenges for detection of a counterpart are the uncertainty in the EM signature of any event, the large positional uncertainty ($\sim$hundreds of square degrees) associated with any trigger, and the large number of candidate transients that are therefore found via a systematic search of the resulting sky region. While the field-of-view of the Liverpool Telescope is such that it cannot compete with modern synoptic survey facilities in the transient search role, it is well equipped to classify transients reported to the collaboration. 

We obtained follow-up observations in response to all three of the triggers, observing a total of $27$ transient sources. A significant number of the transients we classify are unrelated supernovae at distances comparable to those associated with the gravitational wave events. In advance of future observing runs we have made some software enhancements to the telescope to minimise response time during observing programmes of this nature: response time being a key benefit of robotic observing. We will also likely diversify our follow-up approach in future runs, making use of a wider range of the available instrumentation on the telescope. Given the uncertain EM signature of the true counterpart, it is uncertain as to what will be the key diagnostic to set it apart from a long list of unrelated candidates.

\section*{ACKNOWLEDGEMENTS}
The results presented in this paper are based on observations collected with the Liverpool Telescope, which is operated on the island
of La Palma by Liverpool John Moores University in the Spanish
Observatorio del Roque de los Muchachos of the Instituto de Astrofisica de Canarias with financial support from the UK Science
and Technology Facilities Council. This research has made use of NASA's Astrophysics Data System Bibliographic Services; and the SIMBAD data base, operated at CDS, Strasbourg, France. A.S.P. and the development of SPRAT were supported by the EU FP7.2 OPTICON project. C.G.M. acknowledges support from the Royal Society, the Wolfson Foundation and the Science and Technology Facilities Council. We thank the anonymous referee for their useful comments.

\bibliography{ltgwem}

\end{document}